\def\year{2017}\relax
\documentclass[letterpaper]{article} 
\usepackage{aaai17}  
\usepackage{times}  
\usepackage{helvet}  
\usepackage{courier}  
\usepackage{url}  
\usepackage{graphicx}  
\usepackage{amsmath}
\usepackage{amsthm}
\usepackage{amssymb}
\usepackage{subfig}
\usepackage[linesnumbered,ruled]{algorithm2e}

\DeclareMathOperator*{\argmax}{argmax}

\begin{document}
	
	\title{Small Profits and Quick Returns: A Practical Social Welfare Maximizing Incentive Mechanism for Deadline-Sensitive Tasks in Crowdsourcing}
	\author{Duin Back\\
		Department of Computer Science\\
		\small The State University of New York, Korea\\
		\& Stony Brook University, NY, USA\\
		dback@cs.stonybrook.edu \And
		Bong jun Choi\\
		Department of Computer Science\\
		\small The State University of New York, Korea\\
		\& Stony Brook University, NY, USA\\
		bongjun.choi@stonybrook.edu \And
		Jing Chen\\
		Department of Computer Science\\
		StonyBrook University, NY, USA\\
		jchen@cs.stonybrook.edu
	}
	\maketitle
	\begin{abstract}
		As the driving force of crowdsourcing is the interaction among participants, various incentive mechanisms have been proposed to attract sufficient participants. However, the existing works assume that all the providers always meet the deadline and the task value accordingly remains constant. To bridge the gap of such impractical assumption, we model the heterogeneous punctuality behavior of providers and the task value depreciation of requesters. Based on those models, we propose an Expected Social Welfare Maximizing (ESWM) mechanism that aims to maximize the expected social welfare in polynomial time. Simulation results show that our heuristic-based mechanism achieves higher expected social welfare and platform utility via attracting more participants. 
	\end{abstract}
	
\section{Introduction}
	\noindent The existing incentive mechanisms for crowdsourcing assume that all the providers always meet the deadline of requested tasks. In practice, however, we cannot guarantee such perfect punctuality. Therefore, the existing mechanisms, despite their well-defined system models, do not fully reflect the realistic behavior of providers.
	 
	Besides, without guaranteeing the perfect punctuality, there can be task value depreciation if tasks are completed after the deadline. Typically, a requester can achieve a full task valuation if its requested task is completed within the deadline. Otherwise, the later task is completed, the less valuation it achieves \cite{depreciation}. In addition, the rate of depreciation can be different among participants. Therefore, due to such potential task value depreciation, the existing works may not accurately estimate task valuation in crowdsourcing system.
	 
	In addition, such task value depreciation can affect the payment policy in incentive mechanisms. In the existing works, providers are rewarded with a fixed payment policy. However, when tasks are deadline-sensitive, the valuation of tasks depreciates after the deadline or even become valueless, which will inflict loss of utility to requesters. Despite such loss, they are not accounted for their loss. As a result, the satisfaction level of requesters will degrade. 
	
	Therefore, we model the providers' punctuality behavior and the task value depreciation to build an incentive mechanism under more practical and realistic assumptions. Based on our new system model, we propose an expected social welfare maximizing (ESWM) mechanism that aims to maximize the expected social welfare in polynomial time.
	
	\section{Related Works}
	Yang et al. \cite{Yang} presented two models of incentive mechanisms: platform-centric model and user-centric model to motivate mobile users' participation. By rewarding participants proportionally to their contribution, D. Peng et al. \cite{quality} proposed a quality based incentive mechanism for crowdsensing. To maintain sufficient participants and promote dropped users to participate again, Lee and Hoh \cite{Lee} propose a mechanism, called RADP-VPC, to provider long-term incentives to participants. 
	
	\begin{figure*}[!htb]
			\centering
			\subfloat[Social Welfare \label{subfig-2:sw}]{%
				\resizebox{0.235\textwidth}{!}{\includegraphics[width=\linewidth]{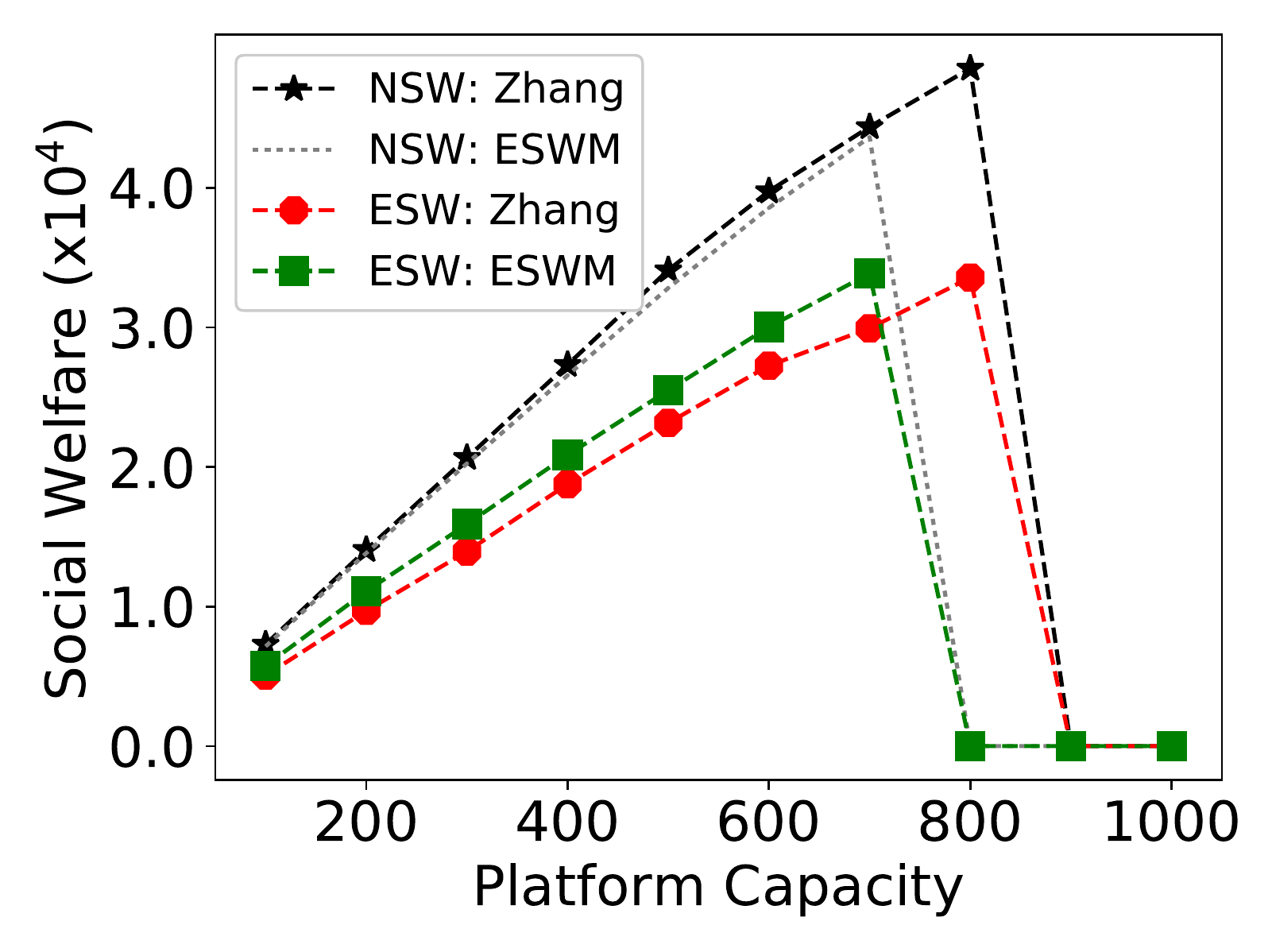}}
			}
			\subfloat[Platform Utility \label{subfig-2:platform}]{%
				\resizebox{0.235\textwidth}{!}{\includegraphics[width=\linewidth]{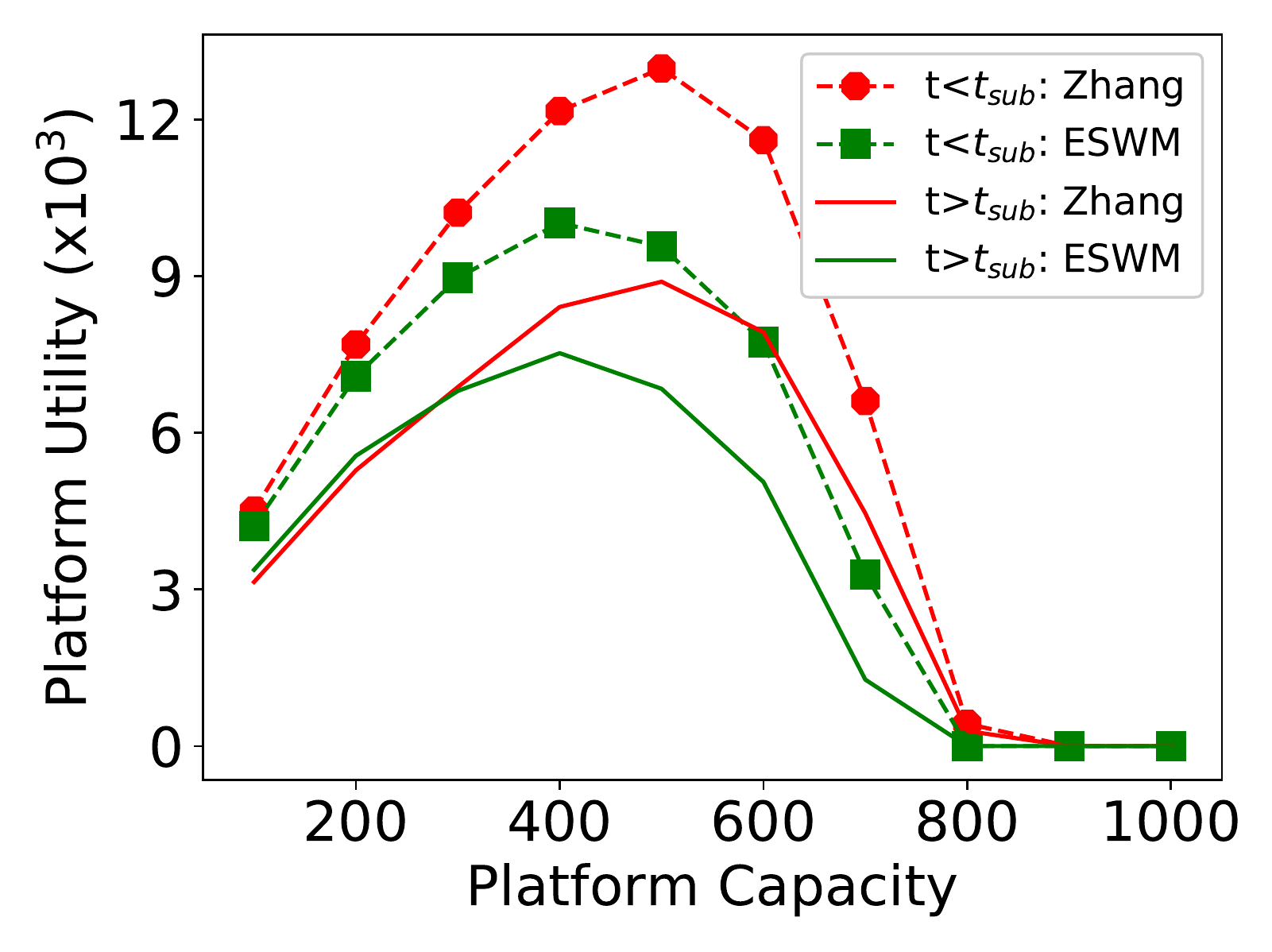}}	}
			\hfill
			\subfloat[Average Requester Utility \label{subfig-2:requester}]{%
				\resizebox{0.235\textwidth}{!}{\includegraphics[width=\linewidth]{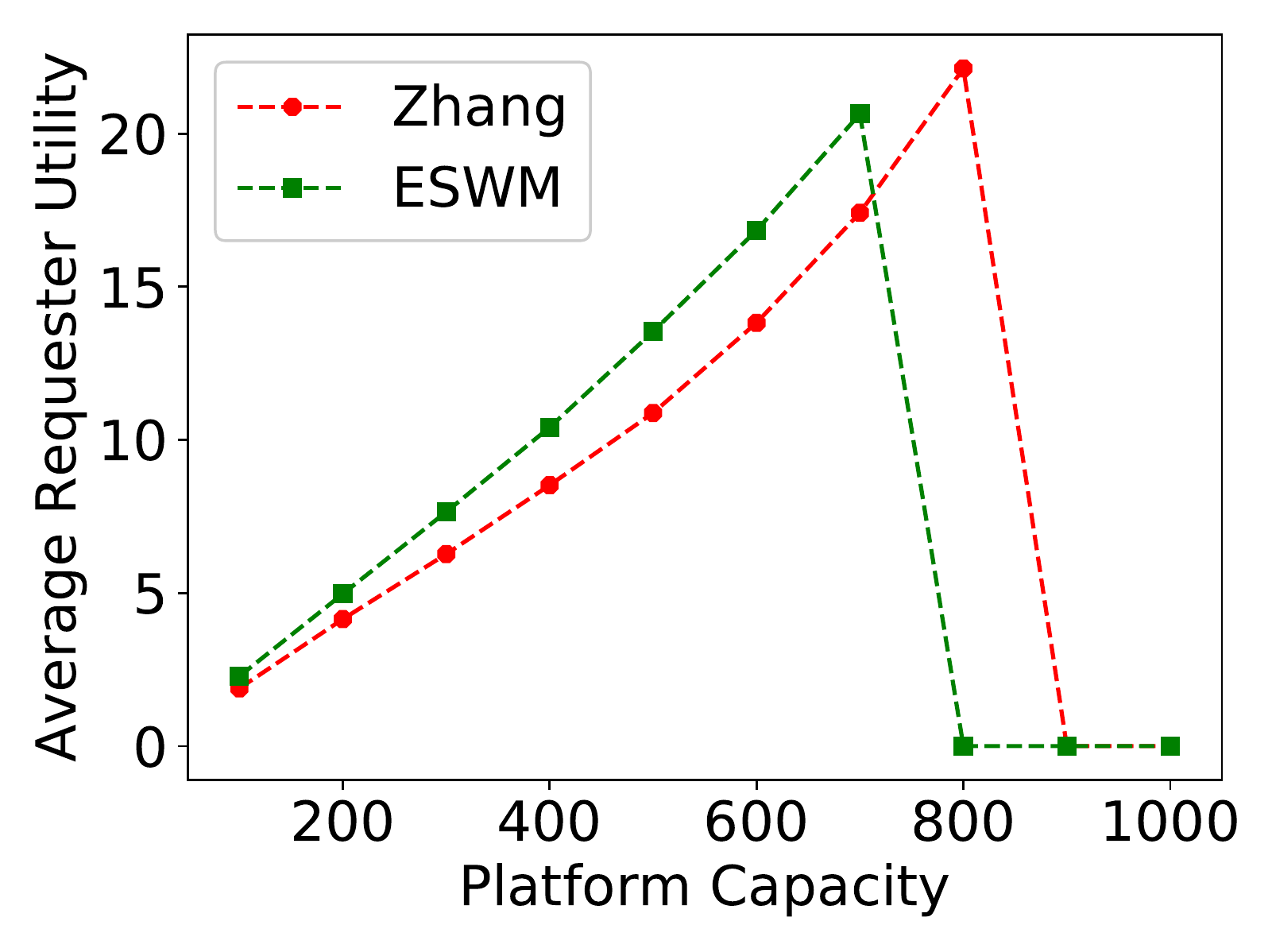}}	}
			\subfloat[Average Provider Utility \label{subfig-2:provider}]{%
				\resizebox{0.235\textwidth}{!}{\includegraphics[width=\linewidth]{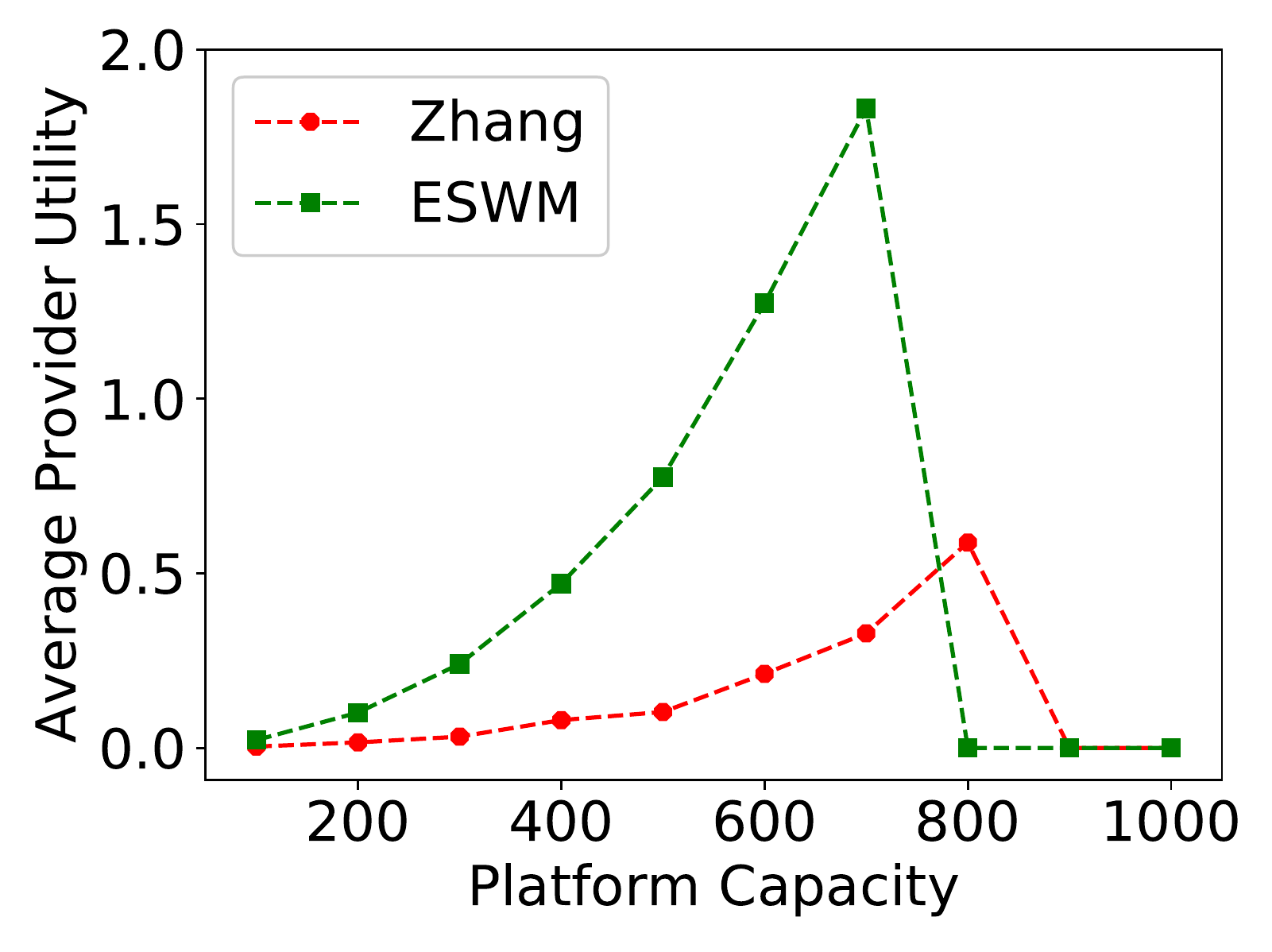}}	}
			\caption{Benchmark VS ESWM}
			\label{fig:5}
	\end{figure*}
	\begin{figure*}[!htb]
		\centering
		\subfloat[Social Welfare \label{subfig-1:sw}]{%
			\resizebox{0.235\textwidth}{!}{\includegraphics[width=\linewidth]{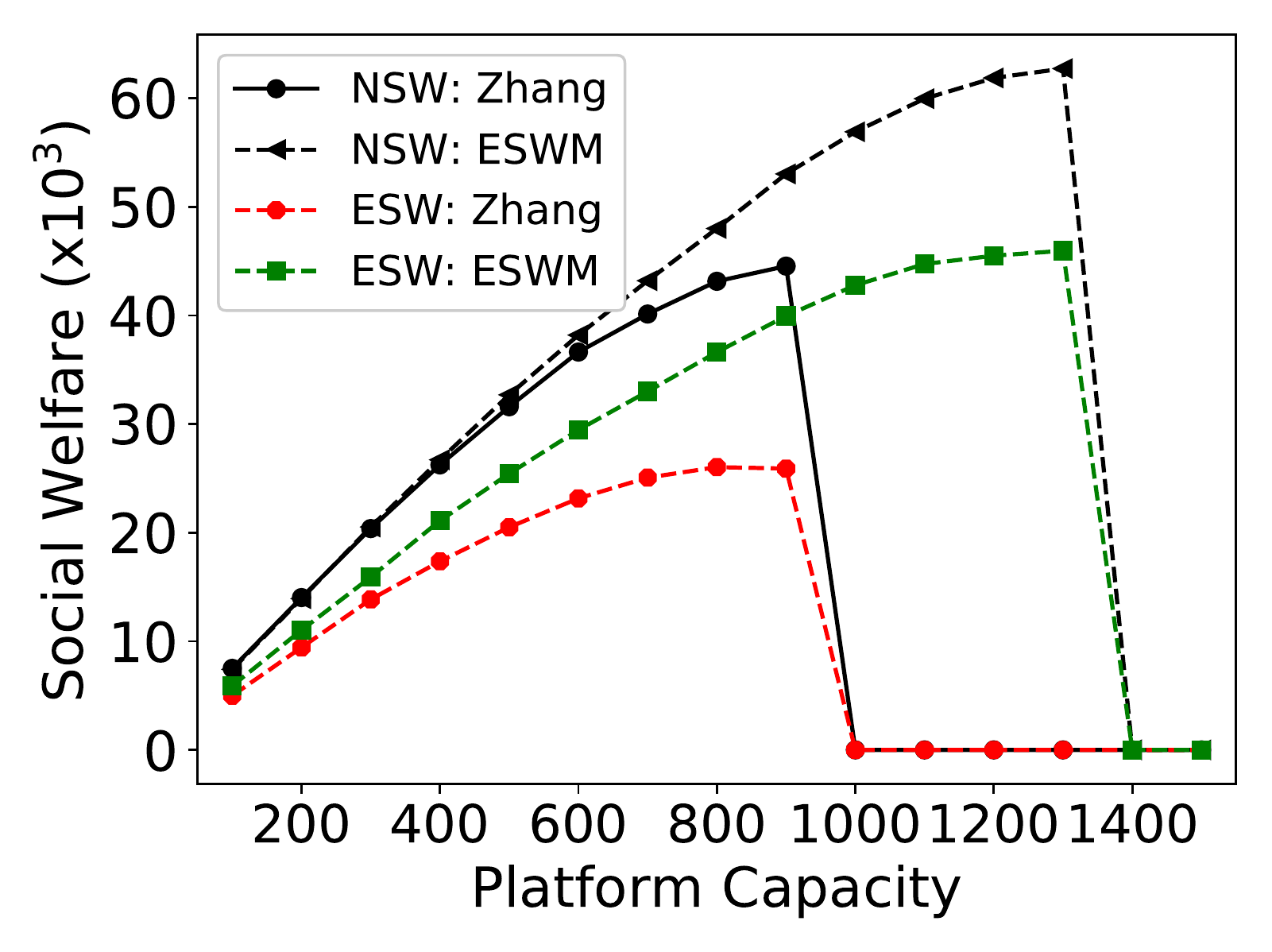}}
		}
		\subfloat[Platform Utility \label{subfig-1:platform}]{%
			\resizebox{0.235\textwidth}{!}{\includegraphics[width=\linewidth]{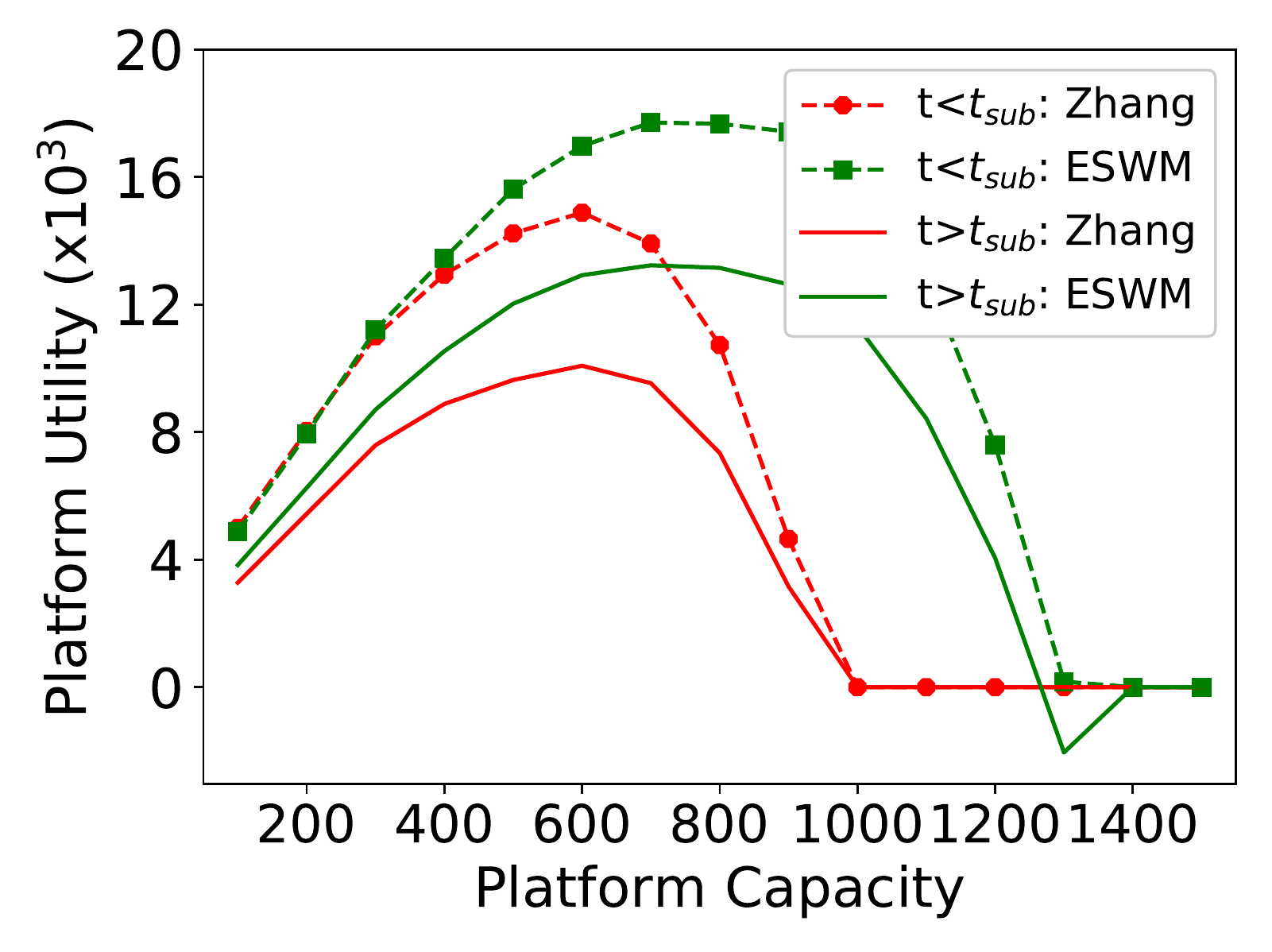}}	}
		\hfill
		\subfloat[Average Requester Utility \label{subfig-1:requester}]{%
			\resizebox{0.235\textwidth}{!}{\includegraphics[width=\linewidth]{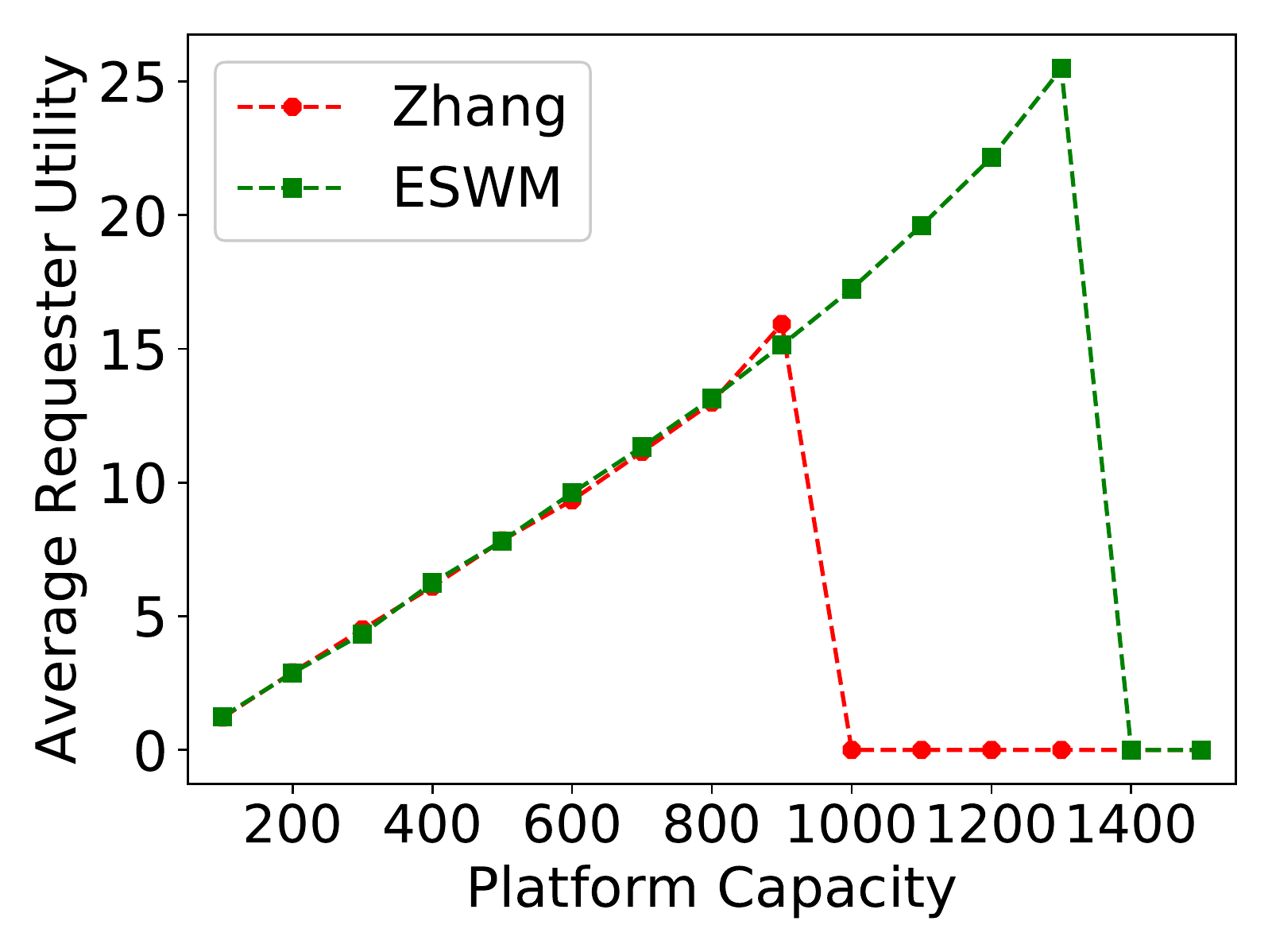}}	}
		\subfloat[Average Provider Utility \label{subfig-1:provider}]{%
			\resizebox{0.235\textwidth}{!}{\includegraphics[width=\linewidth]{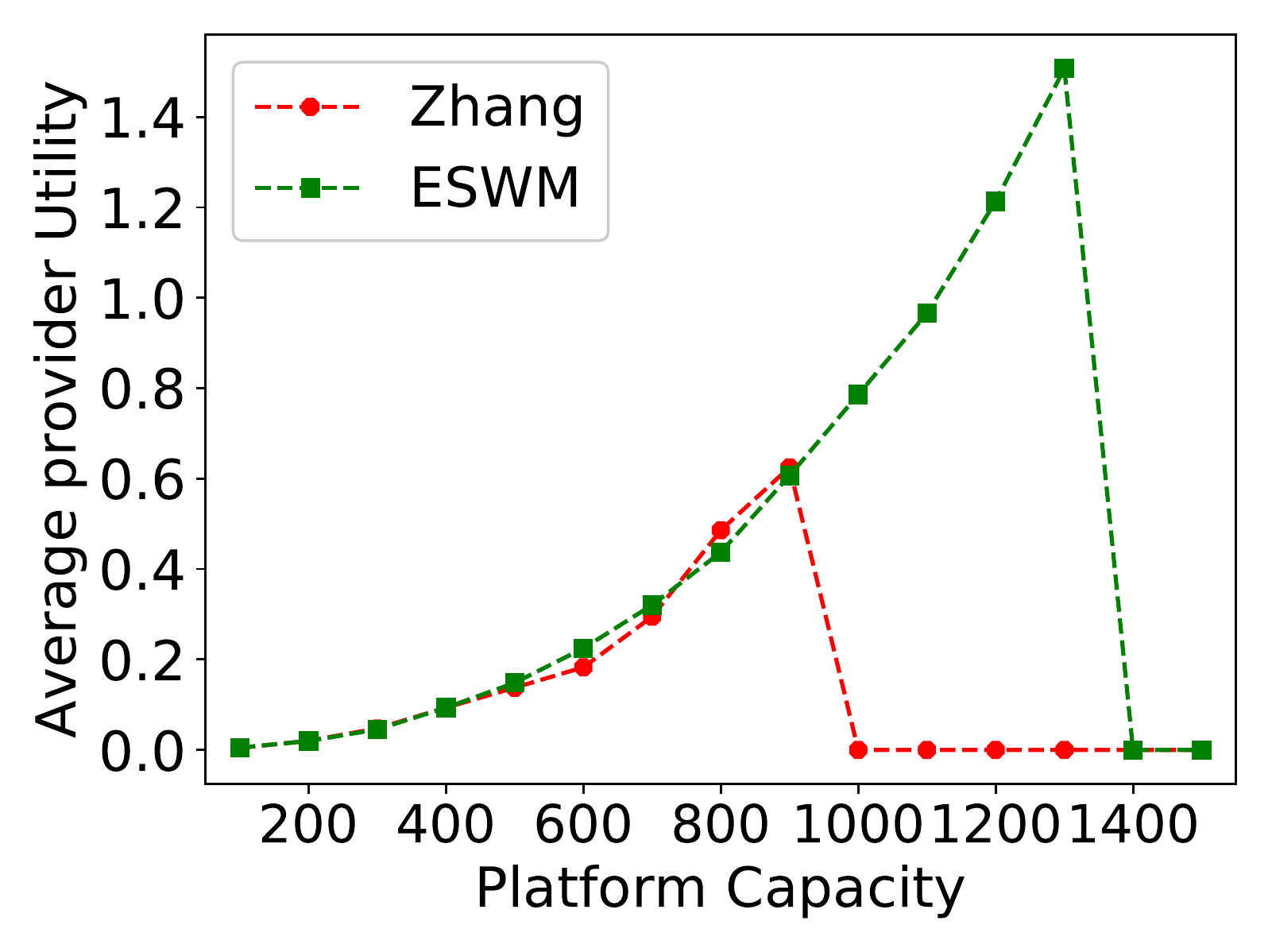}}	}
		\caption{Benchmark VS ESWM with reselection}
		
		\label{fig:6}
	\end{figure*}
		
	\section{Expected Social Welfare Maximizing Problem}
	\label{part:3}
	The objective of the platform is to find the optimal requester-provider matches \small$L$ \normalsize that maximize the expected social welfare as formulated below 
	\small
	\begin{equation}
	\label{eq:15}
	L^*=\argmax_{L} \sum_{r_j \in R}\sum_{w_i \in W} (\mathbf{E}_i(v_j(t)) - c_i) l_{ji},  
	\end{equation}\normalsize
	subject to 
	\small\begin{equation}
	\tag{11.a}
	\label{eq:11}
	\sum_{r_j \in R} x_j \leq K,
	\end{equation}
	\begin{equation}
	\tag{11.b}
	\label{eq:13}
	\sum_{r_j \in R} x_j = \sum_{r_j \in R}\sum_{w_i \in W} l_{ji} = \sum_{w_i \in W} y_i,
	\end{equation}
	\begin{equation}
	\tag{11.c}
	\label{eq:12}
	x_j \in \{0, 1\},\quad \forall r_j \in R,
	\end{equation}
	\begin{equation}
	\tag{11.d}
	\label{eq:14}
	y_i \in \{0, 1\},\quad \forall w_i \in W,
	\end{equation}
	\begin{equation}
	\tag{11.e}
	\label{eq:16}
	l_{ji} \in \{0, 1\},\quad \forall l_{ji} \in L.
	\end{equation}\normalsize
	where \small$r_j$ ($R$) \normalsize and \small$w_i$ ($W$) \normalsize denote a requester (a set of requesters) and a provider (a set of providers), respectively. \small$r_j$ \normalsize wants to complete a task with certain deadline, of which valuation is \small$v_j$ \normalsize. \small$w_i$ \normalsize wants to work on a requested task and get rewarded for the task to compensate the incurred cost, \small$c_i$\normalsize. \small$\mathbf{E}_i(v_j(t))$ \normalsize is the expected task valuation of a task from \small$r_j$ \normalsize when completed by \small$w_i$\normalsize. \small$l_{ji}$ \normalsize indicates that \small$r_j$ \normalsize and \small$w_i$ \normalsize are matched together. $x_j$ and $y_i$ indicate whether $r_j$ and $w_i$ are selected or not. 	
	Constraint (\ref{eq:11}) specifies that the platform has a limited capacity to handle \small$K$ \normalsize task requests and constraint (\ref{eq:13}) specifies that each selected requester will be matched to only one provider. To obtain the optimal solution in (\ref{eq:15}), we need to solve a binary integer programming problem, which is NP-complete. 
	Thus, to overcome such impracticality, we propose an expected social welfare maximizing mechanism (ESWM) that is based on a greedy algorithm to heuristically obtain the locally optimal solution.

	\section{Performance Evaluation}
	\label{part:6}
	As performance metrics, we consider the na\"\i ve social welfare (NSW) to simply sum the task valuation and the expected social welfare (ESW), the platform utility, and the average utility of requesters and providers. We compare the performance metrics of our mechanism to those of the benchmark \cite{team1} whose winner selection process is based on the greedy algorithm, but only considering the platform utility.  
	
	In reality, as all the participants are rationally selfish, they are likely to select the mechanism that provides higher utility to them. Thus, when both the benchmark and our mechanism exist in a crowdsourcing system, the number of participants that each mechanism attracts can vary depending on the average utility each mechanism provides. 
	To reflect such difference of attractiveness between two mechanisms to participants, we set the participation probability of a participant proportional to the square root of the average utility of participants in Figure \ref{fig:5} where both mechanisms are given the same number of requesters and providers, based on \cite{concave}. According to the probabilities, each participant decides which mechanism it will participate in. We call such decision making process  \textit{reselection}. 
	Figure \ref{subfig-1:sw} and Figure \ref{subfig-1:platform} show that the ESWM mechanism achieves higher social welfare and platform utility than the benchmark. Such outperformance can be achieved as the ESWM mechanism can attract more participants, which increase the chance of getting better providers. 
	Figure \ref{subfig-1:requester} and Figure \ref{subfig-1:provider} show that the ESWM mechanism and the benchmark achieve almost the same average utility as long as both can handle task requests. This is because the increased number of requesters and providers participating in the ESWM mechanism ironically decreases the average utility of participants.  
	Based on our observation from Figure \ref{subfig-1:requester} and Figure \ref{subfig-1:provider}, we can anticipate that there will not be a significant increase in the number of reselection again as the benchmark and the ESWM mechanism reached the balance point of the average utility of participants. In addition, the ESWM mechanism can support more task requests. 
		
	\section{Conclusion}
		In this work, we proposed an Expected Social Welfare Maximizing (ESWM) mechanism that is based on a greedy algorithm to heuristically obtain the locally optimum in polynomial time. 
		Simulation results show that the ESWM mechanism achieves higher expected social welfare and platform utility than those of the benchmark mechanism via attracting more participants. 
		
		\section{Acknowledgments}
		This research was supported by the Ministry of Science, ICT and Future Planning (MSIP), Korea, under the “ICT Consilience Creative Program” (reference number IITP-2015-R0346-15-1007) supervised by the Institute for Information and Communications Technology Promotion (IITP).
		
		\fontsize{9.0pt}{10.0pt}
		\selectfont
		\bibliographystyle{aaai}
		\bibliography{RPE}
		
	\end{document}